\documentstyle[aps,pra,epsfig,twocolumn]{revtex}
\begin{document}
\draft
\wideabs{
\title{Stability of standing matter waves in a trap}
\author{A.E. Muryshev${}^1$, H.B. van Linden van den Heuvell${}^2$, and
G.V. Shlyapnikov${}^{1,3,4}$}
\address{${}^1$ Russian Research Center, Kurchatov Institute, Kurchatov
Square, 123182 Moscow, Russia\\
${}^2$ Van der Waals-Zeemann Institute, University of Amsterdam,
Valckenierstraat 65, 1018 XE Amsterdam, The Netherlands\\
${}^3$ FOM Institute for Atomic and Molecular Physics, Kruislaan 407, 1098
SJ Amsterdam, The Netherlands \\
${}^4$ Laboratoire Kastler Brossel $^{*}$, 24, Rue Lhomond, F-75231 Paris
Cedex 05, France}
\date{\today}
\maketitle
\begin{abstract}
We discuss excited Bose-condensed states and find the criterion of
dynamical stability of a kink-wise state, i.e., a standing matter
wave with one nodal plane perpendicular to the
axis of a cylindrical trap. The dynamical stability requires a strong
radial confinement corresponding to the
radial frequency larger than the mean-field
interparticle interaction. We address the question of thermodynamic
instability related to the presence of excitations with negative energy.
\end{abstract}
\pacs{03.75.Fi,05.30.Jp}
}
\narrowtext

The discovery of Bose-Einstein condensation (BEC) in trapped clouds
of alkali atoms \cite{BEC} and extensive studies of Bose-condensed
gases in the recent years have led to the observation of new
macroscopic quantum phenomena, such as interference between two
independently created condensates \cite{interf} and reduction of
the rates of inelastic processes (three-body recombination) in the
presence of a condensate \cite{rates}.
The success in these studies stimulates an interest in
macroscopically excited
Bose-condensed states, i.e., excited-state solutions of the
Gross-Pitaevskii equation, where one can expect to observe novel
signatures of BEC.
A widely discussed example is the vortex state, well known in
superfluid liquid helium \cite{helium}.

Another option concerns excited Bose-condensed states which have a
macroscopic wavefunction with nodal planes perpendicular to the
symmetry axis of a trap.
These states represent standing matter waves for which the trap
serves as a cavity, and it will be worth studying
in which aspects they are similar to light waves.
An interesting idea concerns
an atom laser for the generation of coherent matter waves which,
due to the potential presence of nodal planes in the condensate
wavefunction, will be quite different from the matter waves out of
a ground-state Bose condensate.
The waves with one nodal plane (kinks or dark solitons), being the
lowest energy phase slip, are of great interest in connection with
the decay of persistent currents.
Suggested ways of creating standing matter waves in a trap rely on
adiabatic Raman transfer of particles from the ground to
excited Bose-condensed state \cite{Zoller} or on selective population of
trap levels by bosonically enhanced spontaneous emission of
optically excited atoms of an incoming beam \cite{Ben}.

A principal question concerns the stability of excited Bose-condensed
states with respect to the interparticle interaction. In this paper
we consider standing matter waves with one nodal plane perpendicular to
the axis of a cylindrical trap.
For the axially Thomas-Fermi regime (axial frequency
$\omega_z$ is much smaller than the mean-field interaction) this
state can be called "kink-wise" (see \cite{R} and Fig.1),
as the presence of the nodal plane makes a kink in the dependence of
the condensate wavefunction $\Psi_0$ on the axial coordinate.
A characteristic size of the kink is of order the correlation length,
and the corresponding (axial) kinetic energy of the condensate is of
\begin{figure}[tbp]
\epsfig{file=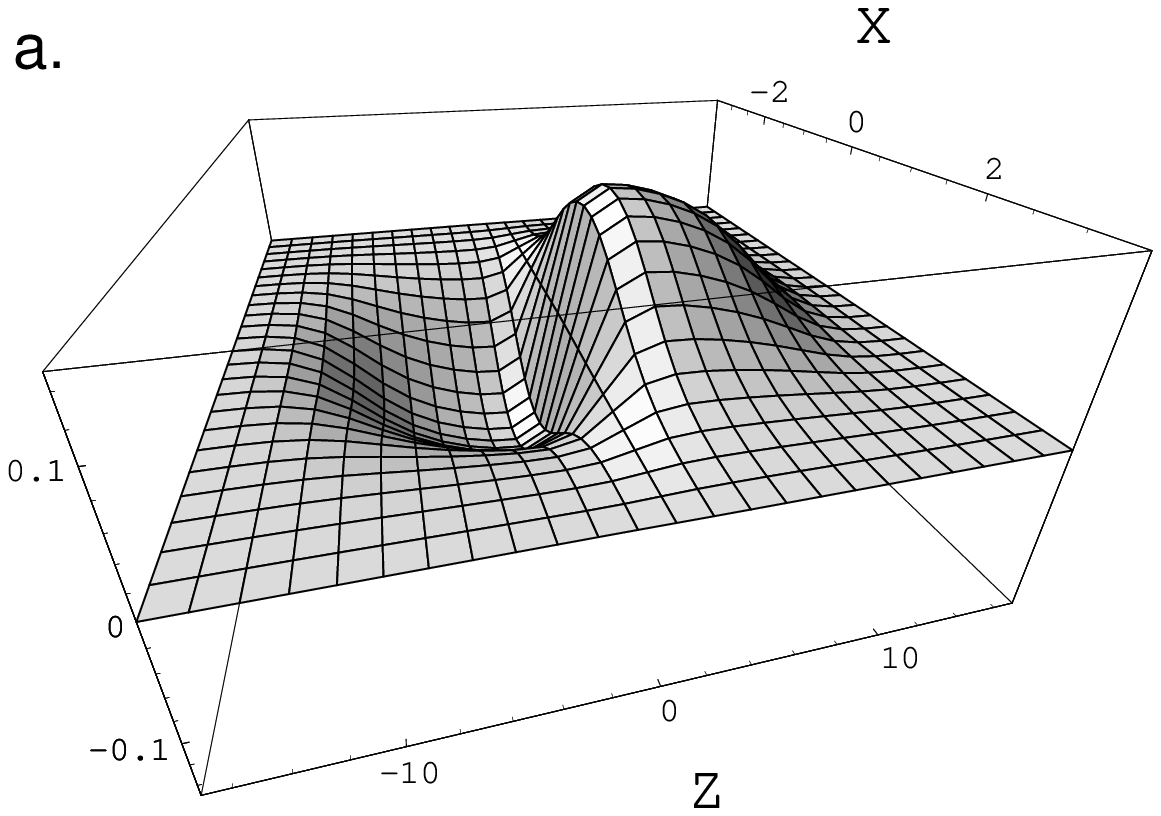, width=0.9\linewidth}
\epsfig{file=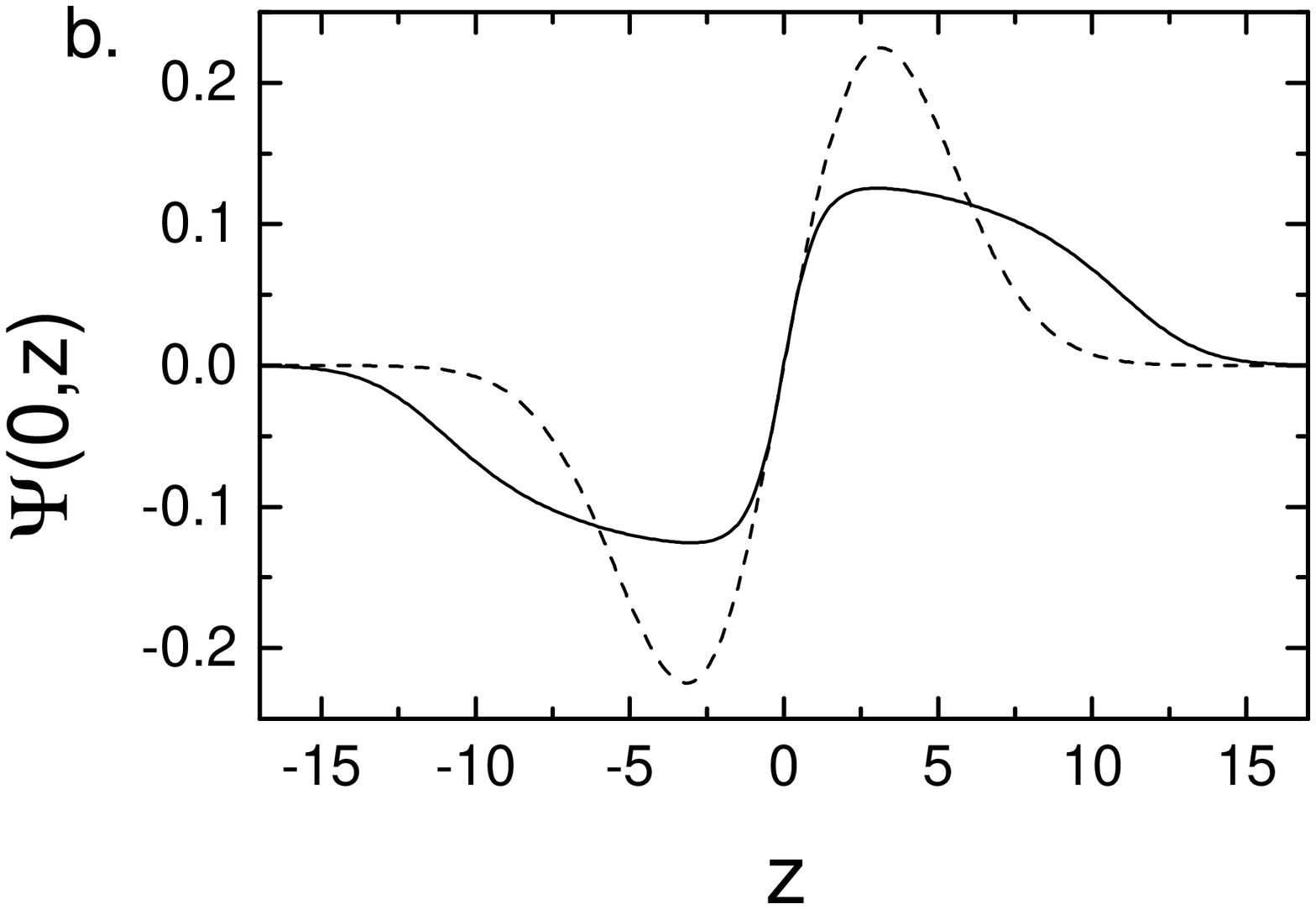, width=0.9\linewidth}
\vspace{2mm}
\caption{ \protect
Condensate wavefunction $\Psi_0$ of a kink-wise state in a cylindrical trap
for the ratio of the mean-field interaction at maximum density ($n_{0m}$)
to the trap frequencies,
$n_{0m}\tilde U:\hbar\omega_{\rho}:\hbar\omega_z=10:10:1$,
($\tilde U=4\pi\hbar^2a/m$, $a>0$ is the scattering length, and $m$
the atom mass).
The axial ($z$) and radial ($x$) coordinates are given
in units of the correlation length $l=\hbar/\sqrt{n_{0m}\tilde U}$,
and (normalized to unity) $\Psi_0$ in units of $l^{-3/2}$.
In (b) the solid curve is $\Psi_0(0,z)$, and the dashed curve
the first (axially) excited state of a harmonic oscillator.
}
\label{1}
\end{figure}
\hspace{-3.5mm}order
the mean-field interaction.
Similarly to the case
of vortices \cite{Rokhsar,Fetter},
the instability of the kink-wise state is
related to the motion of the kink (core) with respect to the rest
of the condensate. We analyze the spectrum of elementary
excitations of this Bose-condensed state and find the criterion of dynamical
stability, i.e., the stability of small-amplitude normal modes:
In order to prevent the interaction-induced transfer of (axial) kinetic
energy of the condensate to the radial degrees of freedom one should strongly
confine the radial motion by making the radial
frequency $\omega_{\rho}$ larger than the mean-field interparticle interaction.
Under this condition the kink-wise state will be perfectly
stable in the limit of zero temperature: The thermodynamic instability
related to the presence of an excitation mode with
negative energy will not lead to decay in the absence of dissipative
processes.

Our conclusion of how tightly one should confine the radial motion to achieve
(quasi)1D dynamics of a condensate and observe dynamically stable kinks
(dark solitons) is directly related to the problem of "engineering the
dimensionality of space". Various experiments aiming for quasi-1D gases are
currently being setup.

To gain insight in the nature of the instability, we first consider a kink-wise
Bose-condensed state in the absence of trapping field, i.e., the state with
one nodal $x,y$ plane in an otherwise spatially homogeneous condensate
of density $n_0$.
We will use the chemical potential $\mu=n_0\tilde U$
and the correlation length $l=\hbar/\sqrt{m\mu}$ as units of energy and
length, and $n_0$ as unit of density. For a positive scattering length
the Gross-Pitaevskii equation is reduced to
\begin{equation}       \label{GPE1}
-\frac{1}{2}\frac{d^2\Psi_0}{dz^2}+\Psi_0^3-\Psi_0=0
\end{equation}
and has a simple solution describing the kink in the dependence of $\Psi_0$
on the $z$ coordinate (see, e.g., \cite{R}):
\begin{equation}    \label{th}
\Psi_0=\tanh{z},
\end{equation}
where $z$ is counted from the nodal plane.
The presence of the kink results in a large kinetic energy $K$ ($\sim \mu$)
per condensate particle at distances of order $l$ from the nodal plane,
which we below call kink-related kinetic energy:
\begin{equation}   \label{K}
K(z)=-\frac{1}{2}\Psi_0\frac{d^2\Psi_0}{dz^2}=
\frac{\tanh^2{z}}{\cosh^2{z}}.
\end{equation}
Elementary excitations of the kink-wise condensate are characterized
by the momentum ${\bf k}$ of transverse ($x,y$) motion and by the
quantum number $\nu$ of motion along the $z$ axis.
In terms of the operators $\hat b_{k\nu},\hat b_{k\nu}^{\dagger}$ of the
excitations the above-condensate part of the field operator of atoms
can be represented as
$\sum_{k\nu}\exp(i{\bf kr})[u_{k\nu}(z)\hat b_{k\nu}-v_{k\nu}(z)\hat
b_{-k\nu}^{\dagger}]$,
and we obtain the following Bogolyubov-de Gennes
equations for the excitation energies $\varepsilon_{k\nu}$ and
wavefunctions $f^{\pm}=u_{k\nu}\pm v_{k\nu}$:
\begin{equation}
\varepsilon_{k\nu}f^{\mp}(z)=[h_{\pm}+(k^2/2)]f^{\pm}(z),   \label{Bog1}  \\
\end{equation}
where the operators $h_+=-d^2/2dz^2+\Psi_0^2-1$ and $h_-=-d^2/2dz^2+
3\Psi_0^2-1$ take the form
\begin{eqnarray}
h_+&=&-\frac{\coth{z}}{2}\frac{d}{dz}\left[\tanh^2{z}\frac{d}{dz}
\coth{z}\right],      \label{h+}\\
h_-&=&-\frac{\cosh^2{z}}{2}\frac{d}{dz}\left[\frac{1}{\cosh^4{z}}
\frac{d}{dz}\cosh^2{z}\right].   \label{h-}
\end{eqnarray}
The kink-wise state has several excitation modes with zero energy.
The ones for $k=0$ are the well-known fundamental modes of a 1D kink state,
following from decoupled equations
$h_{\pm}f^{\pm}=0$. Those which do not exponentially grow at large $z$ are
$f_1^+=\tanh{z},\, f_1^-=0$; $f_2^+=0,\, f_2^-=\cosh^{-2}{z}$, and
$f_3^+=z\tanh{z}-1,\,f_3^-=0$.
The mode $f_2^{\pm}$, which is even with respect to inversion of the
$z$ coordinate, is localized at distances of order $l$ from the nodal plane.

As an example, we demonstrate the instability of transverse normal modes
in the absence of translational motion of the nodal plane.
We consider modes
which in the limit $k\rightarrow 0$ correspond to the localized
zero-energy mode $f_2^{\pm}$.
For transverse momenta $k\ll 1$ the excitation wavefunctions at
distances $z\ll k^{-1}$ can be found as series of expansion in powers of $k$.
As the leading term in the expansion for the function $f^-$ is proportional
to $\cosh^{-2}{z}$, from Eqs.(\ref{Bog1}) one can find that
the leading terms in the expansion for the function $f^+$ should be
$f^+={\rm const}$ and $f^+=f_3^+$. Hence, this expansion takes the form
\begin{equation}      \label{exp1}
f^+=C[1+Bf_3^+(z)+k^2g(z)],
\end{equation}
where $C$ and $B$ are constants.
The equation for the function $g$ follows directly from
Eqs.~(\ref{Bog1})-(\ref{h-}):
\begin{equation}     \label{E}
2\varepsilon_k^2=k^2h_-(2h_+g(z)+Bf_3^+(z)+1)-k^2\cosh^{-2}{z}.
\end{equation}
Using Eq.(\ref{h-}) and performing the integration of Eq.(\ref{E}),
we obtain the relation
$(3\varepsilon_k^2/k^2+1)\cosh^2{z}+4+3f_3^+(z)+6h_+g(z)=0$.
As $g$ should not contain terms exponentially growing with
$z$, we find the dispersion relation
corresponding to imaginary excitation
energies:
\begin{equation}    \label{disp}
\varepsilon_k=ik/\sqrt{3}.
\end{equation}
The instability of transverse normal modes, following from Eq.(\ref{disp}),
originates from the transfer of the (longitudinal) kink-related kinetic
energy $K$ of the condensate to these modes.
As $K\sim\mu$, it can be transferred by the mean-field interaction to modes
with small $k$.

The demonstrated instability and Eq.(\ref{disp}) are similar to those in
the case of "domain walls" \cite{Kuz}.
In Fig.2 we present numerical results for ${\rm Im}\,
\varepsilon_k$ as
a function of $k$.
For small momenta it increases linearly with $k$,
in accordance with Eq.(\ref{disp}), and reaches its maximum
at $\!k\!=\!1/\sqrt{2}$. Further increase of $k$ leads to
decreasing ${\rm Im}\,
\varepsilon_k$ which becomes zero at $k\!\!=\!\!1$. At this critical point we
have one more zero-energy solution of Eqs.(\ref{Bog1}):
\begin{equation}       \label{3c}
\varepsilon_k=0,\,\,\,\, f^+\propto1/\cosh{z},\,\,\,\, f^-=0.
\end{equation}
For $k>1$ the energy of free transverse motion, $k^2/2$,
exceeds the kink-related kinetic energy $K$, and the normal modes are
dynamically stable, with positive $\varepsilon_k$.
\begin{figure}[tbp]
\epsfig{file=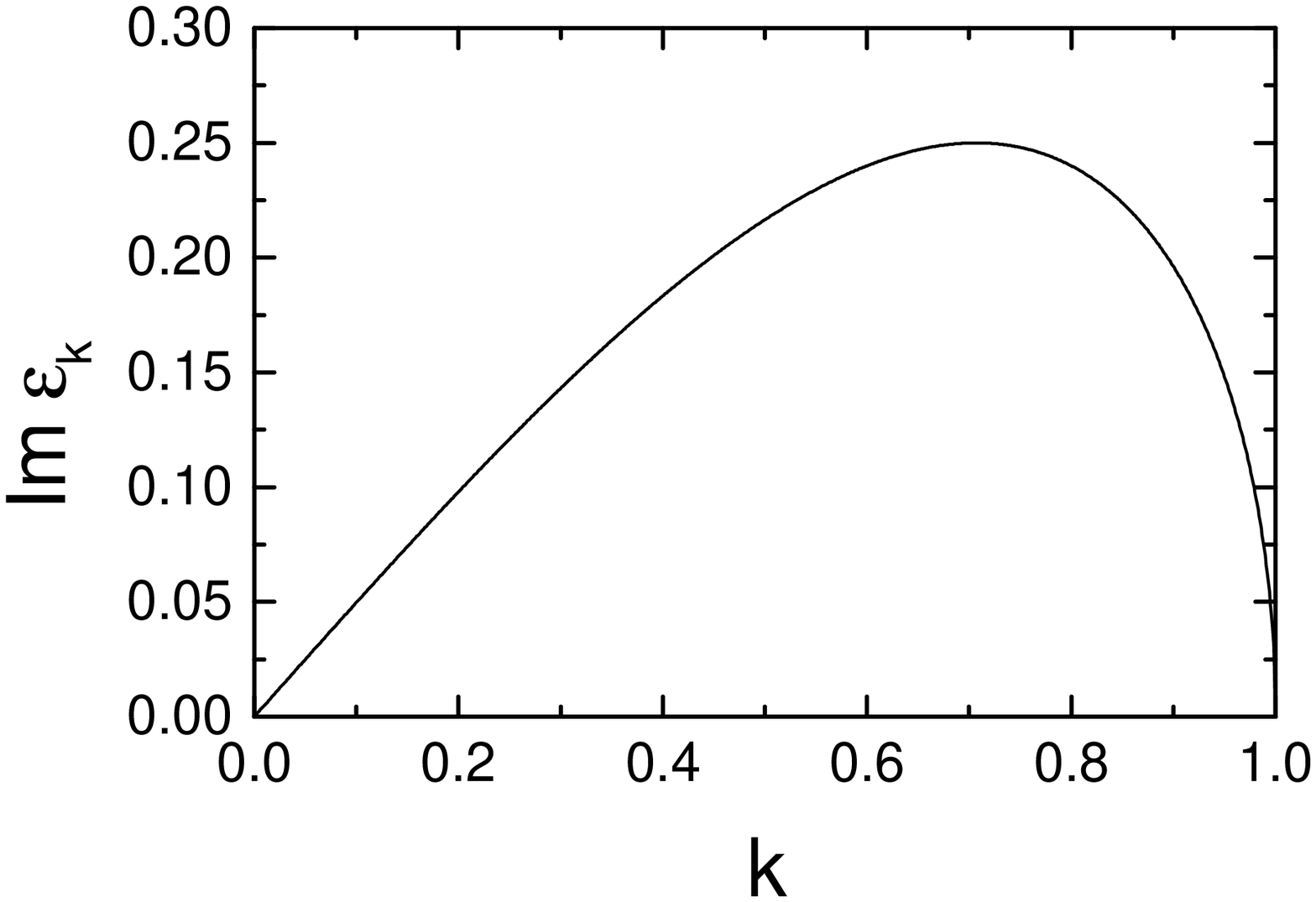, width=0.9\linewidth}
\vspace{2mm}
\caption{ \protect
Imaginary part of the excitation energy (in units of $\mu$) versus the
transverse momentum $k$ (in units of $l^{-1}$) for a
kink-wise condensate in the absence of a trapping field.
}
\label{2}
\end{figure}
One can now understand the origin of
dynamical instability of a kink-wise Bose-condensed state in a
cylindrical harmonic trap: The interparticle interaction
can transfer the (axial) kink-related kinetic energy of the condensate
to the radial degrees
of freedom. In order to suppress this instability one has to
significantly confine the radial motion.
As the (axial) kinetic energy per particle in
the axially Thomas-Fermi condensate is of order the mean-field
interaction at maximum density, $n_{0m}\tilde U$,
the radial frequency should be the same or larger.

We have performed calculations for various
ratia of the radial to axial frequency, $\omega_{\rho}/\omega_z$, and
found the maximum value $\gamma_c$ of the parameter
$\gamma=n_{0m}\tilde U/\hbar\omega_{\rho}$, at which the kink-wise
Bose-condensed state is still dynamically stable, i.e., all
excitation modes have real frequencies. If $\gamma>\gamma_c$,
there are excitations with imaginary
frequencies, and the kink-wise condensate is dynamically unstable.

We have solved the Gross-Pitaevskii equation
\begin{equation}       \label{GPE2}
\!\left[-\frac{\hbar^2}{2m}\Delta+\frac{m}{2}(\omega_z^2z^2+
\omega_{\rho}^2\rho^2)+\tilde U|\Psi_0|^2
-\mu\right]\Psi_0=0,
\end{equation}
together with the Bogolyubov-de Gennes equations for the excitations,
which we write in the form
\begin{eqnarray}
\varepsilon_{\nu}f^\mp&=&\frac{\hbar^2}{2m}\left[ -\Delta +
\frac{\Delta\Psi_0}{\Psi_0}\right] f^\pm + (1 \mp 1)\tilde U|\Psi_0|^2f^\pm.
\label{Bog3}
\end{eqnarray}
Eqs.~(\ref{Bog3})
give real $\varepsilon_{\nu}^2$
which depend continuously on $\gamma$ and the aspect ratio.
In the range of $\gamma$ and $\omega_{\rho}/\omega_z$, where
a given mode $\nu$ is dynamically unstable, $\varepsilon_{\nu}^2<0$
and the energy $\varepsilon_{\nu}$ is purely imaginary.
In the region of dynamical stability $\varepsilon_{\nu}$ is purely real
($\varepsilon_{\nu}^2>0$), and, hence, at the border between the two
regions we have $\varepsilon_{\nu}=0$.

At the critical point $\gamma=\gamma_c$ all excitation energies
$\varepsilon_{\nu}$ are real, and one of the excitations has zero energy.
This is just the mode which for $\gamma>\gamma_c$ becomes
dynamically unstable.
Similarly to the mode of Eq.(\ref{3c}) in the absence of trapping field,
this mode is even with respect to inversion of the $z$ coordinate.
The function $f^-=0$, and $f^+$ follows directly from Eqs.(\ref{Bog3}):
\begin{equation}     \label{crit}
(-\Delta +\Delta\Psi_0/\Psi_0)f^+=0.
\end{equation}
Eq.(\ref{crit}) is the Schr\"odinger equation for the motion of a particle
(with zero energy) in a cylindrically symmetric potential
$V=\hbar^2\Delta\Psi_0/2m\Psi_0$.
The potential $V$
depends on $\gamma$ and the aspect ratio.
Thus, for a given ratio $\omega_{\rho}/\omega_z$ one finds the critical value
$\gamma_c$
by selecting the parameter $\gamma$ such that there is an even (non-zero)
solution of
Eq.(\ref{crit}), remaining finite at the origin and tending to zero at
infinity.
This was checked numerically on the basis of
Eqs.~(\ref{GPE2})-(\ref{Bog3})
for a wide range of $\gamma$ and the aspect ratio.

As follows from our calculations, $\gamma_c$ is minimal for
excitations with the projection of the orbital angular momentum on the
symmetry axis, $M=1$. The dependence of $\gamma_c$ on the aspect ratio
is presented in Fig.3. For $\omega_{\rho}<\omega_z$ even an arbitrary small
interparticle interaction leads to instability, as the axial
"kink-related" energy per particle in the condensate ($\hbar\omega_z$)
can be always transferred to the radial
mode with $M=1$, which by itself has energy $\hbar\omega_{\rho}$.
For $\omega_{\rho}>\omega_z$
the critical value $\gamma_c$ increases with the ratio
$\omega_{\rho}/\omega_z$ and reaches $\gamma_c\approx 2.4$ for
$\omega_{\rho}\gg\omega_z$.
We also found that the decay of dynamically unstable kink states
is accompanied by the undulation of the nodal plane and the formation
of vortex-antivortex pairs, similar to the decay of dark optical
solitons \cite{Kiv}.
\begin{figure}[tbp]
\epsfig{file=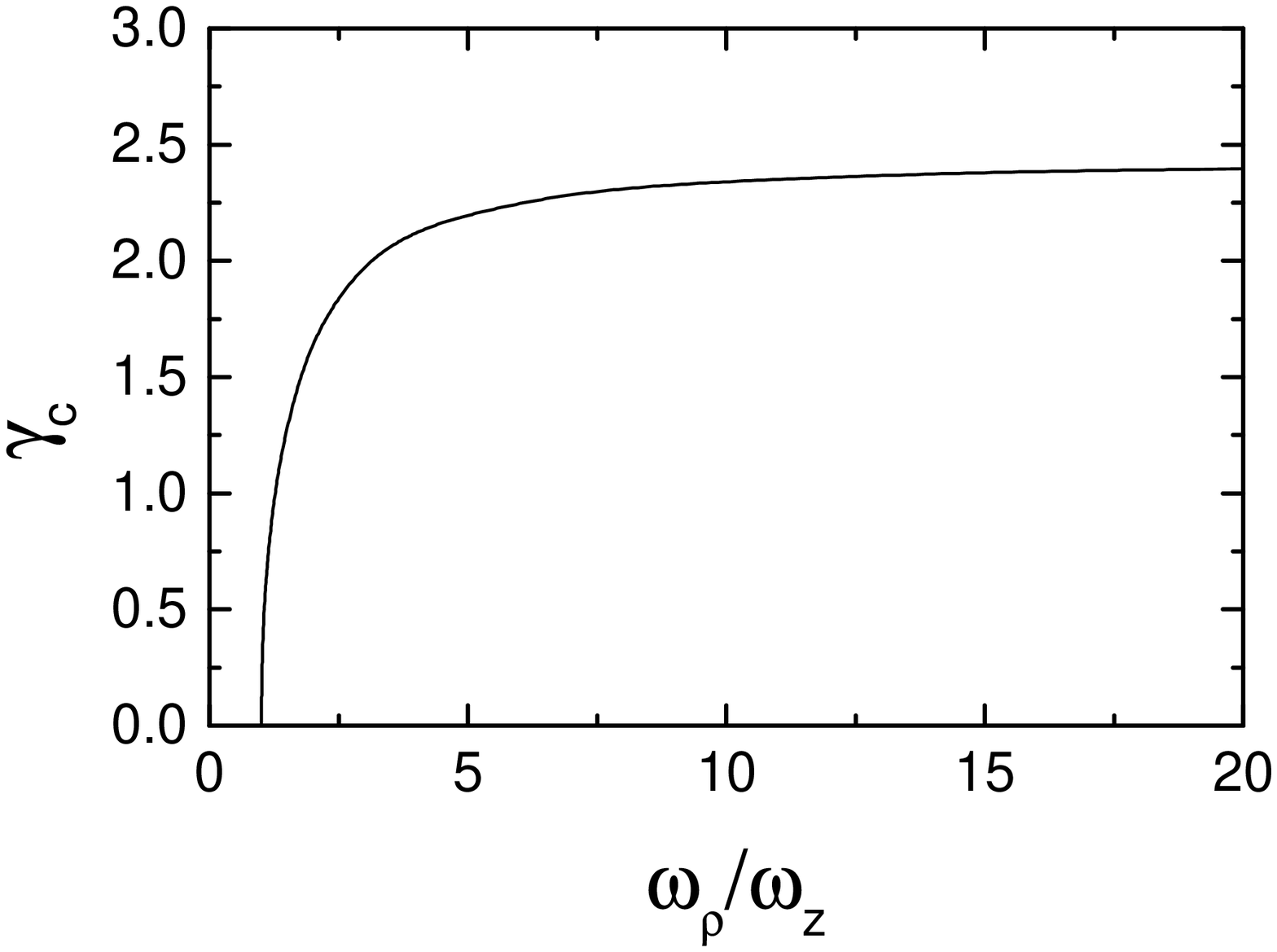, width=0.9\linewidth}
\vspace{2mm}
\caption{ \protect
Critical parameter $\gamma_c$ versus the aspect ratio for a
kink-wise condensate in a cylindrical trap.
}
\label{3}
\end{figure}
The criterion of dynamical stability of a kink-wise condensate,
$\gamma<\gamma_c$, can be satisfied in the conditions of current
BEC experiments. For a rubidium condensate in a cylindrical trap with
$\omega_{\rho}\sim 200\,{\rm Hz}\gg\omega_z$ it requires the maximum
density $n_{0m}\alt 10^{14}$ cm$^{-3}$.

Although for $\gamma<\gamma_c$ the kink-wise condensate is dynamically
stable, there is a thermodynamic instability related to the presence of an
excitation with negative energy.
For a very strong radial confinement
of the axially Thomas-Fermi kink-wise condensate ($\hbar\omega_{\rho}
\gg n_{0m}\tilde U\gg\hbar\omega_z$; $\gamma\ll\gamma_c$) we
calculate a negative
excitation energy
close to $\varepsilon_{*}=-\hbar\omega_z/\sqrt{2}$ characteristic
for the 1-D Thomas-Fermi kink-wise condensate in a harmonic trap.

In the 1-D case we calculate the negative excitation energy analytically,
by solving the Bogolyubov-de Gennes equations at distances $z$ from the
origin, much smaller than the
Thomas-Fermi size of the condensate $R=(2\mu/m\omega_z^2)^{1/2}$.
We represent $\Psi_0$ and the excitation wavefunctions
as series of expansion in powers of small parameter
$\zeta=\hbar\omega_z/\mu$. Then, in the same dimensionless units
as in the absence of trapping field,
the Gross-Pitaevskii equation is given by Eq.(\ref{GPE1}), with an
extra term $\zeta^2z^2\Psi_0/2$ in the left-hand side.
Confining ourselves to the expansion up to $\zeta^2$,
we obtain
\begin{equation}      \label{Psi1}
\Psi_0=\tanh z+\zeta^2\eta(z),
\end{equation}
where the function $\eta(z)$ is determined by the equation
\begin{equation}     \label{zeta}
h_-\eta(z)+(z^2/2)\tanh z=0,
\end{equation}
and is not given because of its complexity. For $|z|\gg 1$
we have $\eta=-{\rm sign}z(1+2z^2)/8$, and Eq.(\ref{Psi1})
recovers the Thomas-Fermi result at $ |z|\ll R/l$.
The Bogolyubov-de Gennes equations take the form
\begin{eqnarray}
\varepsilon_{\nu}f^\mp&=&h_\pm f^\pm+\zeta^2[z^2/2+(4 \mp 2)\eta(z)\tanh z]f^\pm
\label{Bog5}
\end{eqnarray}
Just as in the absence of trapping field, we consider a mode for
which the leading term in
the expansion for the function $f^-$ is proportional to
$1/\cosh^2z$.
For finding the excitation energy it is sufficient to keep the terms
independent of $\zeta$ and proportional to $\zeta^2$ in the expansion
for the function $f^+$.
Then, similarly to Eq.(\ref{exp1}), we obtain
$ f^+ \propto [1+\zeta^2G(z)] $.
The equation for
$G(z)$
follows from Eqs.~(\ref{Bog5}),
and by using Eqs.~(\ref{h-}),(\ref{zeta}) is transformed to
\begin{eqnarray}
& &\left(\frac{\varepsilon_{*}^2}{\zeta^2}-\frac{1}{2}\right)
=h_-[h_+G(z)+z^2]- \nonumber\\
&-&2\cosh^2{z}\frac{d}{dz}\left[\frac{\tanh{z}}{\cosh^4{z}}
\frac{d}{dz}(\cosh^2{z}\eta(z))\right]. \label{eps}
\end{eqnarray}
Integration of Eq.(\ref{eps}) gives at large $|z|$ the relation
$d^2G/dz^2=(\varepsilon_{*}^2/\zeta^2-1/2)\cosh^2{z}
-1$ and, as
$G$ should not contain exponentially growing terms, we obtain
$\varepsilon_{*}^2=\zeta^2/2$. The normalization condition
$\int dzf^+f^-=1$ allows then to conclude that the excitation energy
is negative and, hence, equal to $-\hbar\omega_z/\sqrt{2}$.
The frequency $\omega_z/\sqrt{2}$ for the oscillations of the kink
in a 1-D Thomas-Fermi condensate has also been found in the recent
work \cite{Anglin} from the equation of motion for the kink.

The excitation spectrum for $\varepsilon_{\nu}>0$ follows from the
solution of the Bogolubov-de Gennes equations at large $|z|$, where
the Thomas-Fermi shape of $|\Psi_0|$ is not influenced by the kink.
Then, along the lines of the theory for
the 3D case \cite{Stringari,Ohberg,Szep}, we obtain
a discrete spectrum
$\varepsilon_{\nu}\!=\!\hbar\omega_z\sqrt{\nu(\nu+1)/2}$,
where $\nu$ is a positive integer.

Finally, we analyze the influence of the excitation with negative
energy on the stability of the kink-wise condensate.
Beyond the Bogolubov-de Gennes approach, there is a small coupling of this
excitation to the excitations with positive energies.
But at temperatures $T\rightarrow 0$ there will be no real decay
processes. Those require a simultaneous creation of excitations with
positive and negative energies, with the total excitation energy equal
to zero.
Due to the structure of the discrete spectrum for $\varepsilon_{\nu}>0$,
this conservation
of energy can not be satisfied while
creating a moderate number of excitations with the above found negative
energy $-\hbar\omega_z/\sqrt{2}$.

Thus, under the condition of dynamical stability the kink-wise
Bose-condensed state is perfectly stable at $T\rightarrow 0$.
The decay mechanism in the presence
of a thermal cloud is in some sense similar to that
of temperature-dependent damping
of excitations in trapped Bose-condensed gases
(see \cite{FSW}) and originates from the scattering of thermal
particles on the kink. Accordingly, the decay time can be
made large by decreasing temperature well below the value of the mean-field
interparticle interaction. A detailed analysis of dissipative dynamics
of a kink state at finite $T$ requires a separate investigation.

We acknowledge fruitful discussions with J. Dalibard and Y. Castin.
This work was supported by the Stichting voor Fundamenteel
Onderzoek der Materie (FOM), by INTAS, and by the
Russian Foundation for Basic Studies.
H.B.v.L.v.d.H. acknowledges the hospitality of the RRC Kurchatov Institute,
where part of this work was done.

$^{*}$ L.K.B. is an unit\'e de recherche de l'Ecole Normale Sup\'erieure
et de l'Universit\'e Pierre et Marie Curie, associ\'ee au CNRS.

\end{document}